\documentclass[reprint,amsmath,amssymb,aps]{revtex4}

\usepackage{tensor}
\usepackage{hyperref}
\newcommand{\pfrac}[2]{\frac{\partial{#1}}{\partial{#2}}}
\newcommand{\ppfrac}[3]{\frac{\partial^{2}{#1}}{\partial{#2}\partial{#3}}}

\newcommand{\afffias}{Frankfurt Institute for Advanced Studies (FIAS), Ruth-Moufang-Strasse~1, 60438 Frankfurt am Main, Germany}
\newcommand{\affgu}{Goethe University, Max-von-Laue-Strasse 1, 60438 Frankfurt am Main}

\begin{document}

\preprint{APS/123-QED}

\title{Regularization of the Poisson Bracket in Field Theory}

\author{P. Liebrich}
\affiliation{\afffias}
\affiliation{\affgu}

\date{\today}

\begin{abstract}
In field theory the Poisson bracket $\{F, \mathcal{H}\}$ between an arbitrary function $F$ and the system Hamiltonian $\mathcal{H}$ acquires odd contributions. Here a modification is worked out to remove those terms, which leads to a certain version of the energy-momentum tensor. The resulting Poisson bracket is thus more easily accessible for quantization. An outlook hereupon is given, based on a field theoretic alternative to non-commutative spacetime.
\end{abstract}

\maketitle

\section{Introduction}
The Poisson bracket in mechanics is easily built via anti-symmetric derivation with respect to the canonical coordinates. In multisymplectic field theory there are now $m$ canonical momenta $\pi^\mu$ corresponding to one field variable $\phi$. So the Poisson bracket between functions $F$ and $G$ on phase space is now a vector or a 1-form with components
\begin{equation}\label{PoissonKl}
\{F, G\}_{\phi,\pi^\mu} := \pfrac{F}{\phi} \pfrac{G}{\pi^\mu} - \pfrac{F}{\pi^\mu} \pfrac{G}{\phi}.
\end{equation}
See e.~g. \cite{Redelbach} for a review. This is a serious obstacle to \textbf{quantization}, which fundamentally builds on the translation to the commutator
\begin{equation}
\{F, G\} \mapsto [\hat{F}, \hat{G}].
\end{equation}
On the left-hand side there is now a multi-dimensional object, whereas on the right-hand side the commutator is always a single-dimensional object. This property is compulsory for the quantum commuator: It can be shown that a formal Poisson bracket, i.~e. fulfilling the algebraic properties of a Poisson algebra, between non-commuting objects $\hat{F},\hat{G}$ can only be written in the form
\begin{equation}
\{\hat{F},\hat{G}\} = c (\hat{F}\hat{G} - \hat{G}\hat{F})
\end{equation}
with a commuting coefficient $c$ (``$c$-number''), see e.~g. \cite[section IV.21]{DiracQuant}. It is one of the main tasks in field quantization to solve this impediment. There are various approaches to modify the Poisson bracket to make it suit such a mapping, see e.~g. \cite{Kanatchikov, KanatchikovQuant, Paufler, HippelWohlfarth}.

But already on the classical level there is a further peculiarity with the Poisson bracket when considering brackets with the Hamiltonian density $\mathcal{H}$. When speaking of a ``Hamiltonian'', it is always meant in the covariant sense as a \textbf{De~Donder-Weyl Hamiltonian} on the multisymplectic phase space. The bracket of $\mathcal{H}$ with an arbitrary function $F$ reads
\begin{equation}\label{PoissonDeviation}
\{F,\mathcal{H}\}_\mu = \pfrac{F}{x^\mu} - \left.\pfrac{F}{x^\mu}\right|_\text{expl} + \pfrac{F}{\pi^\mu} \pfrac{\pi^\nu}{x^\nu} - \pfrac{F}{\pi^\nu} \pfrac{\pi^\nu}{x^\mu}.
\end{equation}
(See the appendix for an explicit calculation.) The desirable part is solely the first term, so that this Poisson bracket could be interpreted as the differential of that function $F$. The first correction is already known from mechanics whenever $F$ also depends on the parameter of the system explicitly. The severe part is the remaining terms. They are a deviation in the momentum part and relate to a general gauge freedom in the momenta. It may be considered a kind of ``connection'' to ensure the bracket to lie on the right path. For every representation of the momenta this correction also changes and it has to so that the overall result is balanced again. It can also be seen that in dimension $m = 1$ this deviation vanishes, so it is a pure effect of the multi-dimensional base space.

This paper will be organized as follows: Firstly, the momentum term in eq. (\ref{PoissonDeviation}) will be eliminated via the transition from the Hamiltonian $\mathcal{H}$ to a more suited quantity $\Theta\indices{_\nu^\mu}$. Then the remaining term with the explicit spacetime dependence will be discussed away by means of an extended Poisson bracket. The article concludes with an outlook on quantization of the thus constructed entities.

\section{Correcting the covariant Poisson bracket}
The first aim is to get rid of the momentum term in (\ref{PoissonDeviation}). It has its origin in a general Hamiltonian gauge freedom for the momenta. The canonical equations
\begin{equation}\label{KanGl}
\pfrac{\mathcal{H}}{\pi^\mu} = \pfrac{\phi}{x^\mu} \quad \text{and} \quad \pfrac{\mathcal{H}}{\phi} = -\pfrac{\pi^\mu}{x^\mu}
\end{equation}
only determine the divergence of the covariant momenta, but not the concrete form. What is fixed for the physical system, is the action. The Lagrangian / Hamiltonian is only the integrand, hence different representations can correspond to the same physical system. The equivalence classes are determined by canonical transformations. This freedom shall now be exploited.

The multisymplectic Hamiltonian cannot be interpreted as the energy density, what would be the analogous quantity to the mechanics Hamiltonian. This role is rather taken over by the (00-component of the) \textbf{energy-momentum tensor}
\begin{equation}\label{EnImpTensor}
\theta\indices{_\nu^\mu} := \pfrac{\mathcal{L}}{\pfrac{\phi}{x^\mu}} \pfrac{\phi}{x^\nu} - \delta^\mu_\nu \mathcal{L} = \pi^\mu \pfrac{\mathcal{H}}{\pi^\nu} + \delta^\mu_\nu \left(\mathcal{H} - \pi^\beta \pfrac{\mathcal{H}}{\pi^\beta}\right).
\end{equation}
It may be the more fundamental quantity in field theory; its entries and integrals thereof appear in core formulas of quantum field theory and it is itself the source of the gravitational field equation. The entry $\theta_{00}$ denotes the energy density, the entries $\theta_{0i}$ the momentum density and the spatial entries $\theta_{ij}$ ($i,j=1,2,3$) the momentum flux.

In the same way the canonical equations only determine the divergence of the momenta, they also only determine the divergence of this tensor:
\begin{equation}\label{DivEnImp}
\pfrac{\theta\indices{_\nu^\mu}}{x^\mu} = \left.\pfrac{\mathcal{H}}{x^\nu}\right|_\text{expl},
\end{equation}
as is shown in the appendix. Now define the \textbf{alternative tensor}
\begin{equation}\label{ModTheta}
\Theta\indices{_\nu^\mu} := \theta\indices{_\nu^\mu} - \pfrac{}{x^\beta} \left(\phi \left(\delta_\nu^\beta \pi^\mu - \delta_\nu^\mu \pi^\beta\right)\right) = \delta_\nu^\mu \mathcal{H} - \phi \left(\pfrac{\pi^\mu}{x^\nu} - \delta_\nu^\mu \pfrac{\pi^\beta}{x^\beta}\right).
\end{equation}
The divergence of the additional term can be computed to vanish, see the appendix, thus this tensor constitutes an equivalent representation of the energy-momentum tensor. Of course, its absolute value is different, but in most cases one is anyway only interested in relative values. Therefore, it is not uncommon to change the representation of the energy-momentum tensor, compare e.~g. the Belinfante-Rosenfeld symmetrization. An exception from this rule is the Einstein field equation, which will appear later. It now gives rise to the equations
\begin{equation}\label{ThetaKanGl}
\pfrac{\Theta\indices{_\nu^\mu}}{\phi} = -\pfrac{\pi^\mu}{x^\nu} \quad \text{and} \quad \pfrac{\Theta\indices{_\nu^\mu}}{\pi^\mu} = \pfrac{\phi}{x^\nu}.
\end{equation}
So it really determines all partial derivatives of the momenta. It is clear that the Hamiltonian itself could not provide that complete information because it is only a scalar density, so has too few degrees of freedom for field theory. Moreover, this tensor now fulfills the desired relation
\begin{equation}\label{ThetaPois}
\begin{aligned}
\{F, \Theta\indices{_\nu^\mu}\}_\mu &= \pfrac{F}{\phi} \pfrac{\Theta\indices{_\nu^\mu}}{\pi^\mu} - \pfrac{F}{\pi^\mu} \pfrac{\Theta\indices{_\nu^\mu}}{\phi} \\
&= \pfrac{F}{\phi} \pfrac{\phi}{x^\nu} + \pfrac{F}{\pi^\mu} \pfrac{\pi^\mu}{x^\nu} \\
&= \pfrac{F}{x^\nu} - \left.\pfrac{F}{x^\nu}\right|_\text{expl}.
\end{aligned}
\end{equation}
The tensor itself has 2 open indices, of which 1 is contracted with the Poisson bracket index. In view of quantization, this is pleasant as the vectorial character of the Poisson bracket has to be contracted somehow anyway (usually with a restriction on the temporal component). Nonetheless, the remaining index of the tensor field still leaves this relation a vector / 1-form.

\section{Extended Poisson brackets}
\subsection{Extended Lagrange-Hamilton formalism}
So far the Poisson bracket relation (\ref{ThetaPois}) has been brought to a form analogous to point mechanics. However, one can still go a step forward to also eradicate the last term with the explicit spacetime derivative. The method here can already be used in mechanics, namely \textbf{reparametrizing} the system \cite{GRextended, ParamField}. Then spacetime will appear as a fiber coordinate, hence the additional partial derivative will be included automatically.

Firstly, the Lagrangian density is being reparametrized from the coordinates $x$ in terms of $X$ by means of multiplying by a Jacobian determinant:
\begin{equation}
\tilde{\mathcal{L}}_e := \mathcal{L} \det \pfrac{x}{X}.
\end{equation}
The determinant signifies a change of the integration variables in the integral as the action shall remain untouched. Such a reparametrization further gives rise to the \textbf{additional canonical momenta}
\begin{equation}\label{RaumzeitImpuls}
\tilde{t}\indices{_\mu^\nu} := -\pfrac{\tilde{\mathcal{L}}_e}{\pfrac{x^\mu}{X^\nu}}
\end{equation}
as counterparts to the field variables $x^\mu$. The sign is a convention to draw a connection to the energy from extended mechanics \cite{ExtPointMechanics} as the conjugate to time. The \textbf{extended Hamiltonian density} is now given by
\begin{equation}
\tilde{\mathcal{H}}_e\left(\phi, \tilde{\pi}^\nu, x^\mu, \tilde{t}\indices{_\mu^\nu}, X^\nu\right) := \tilde{\pi}^\nu \pfrac{\phi}{X^\nu} - \tilde{t}\indices{_\mu^\nu} \pfrac{x^\mu}{X^\nu} - \tilde{\mathcal{L}}_e\left(\phi, \pfrac{\phi}{X^\nu}, x^\mu, \pfrac{x^\mu}{X^\nu}, X^\nu\right).
\end{equation}
The tilde above all momentum quantities (and the Lagrangian / Hamiltonian) denotes the additional functional determinant compared to the simply covariant formulation, explicitly:
\begin{equation}
\tilde{\pi}^\nu := \pi^\mu \pfrac{X^\nu}{x^\mu} \det \pfrac{x}{X}.
\end{equation}
The additional canonical variable pair yields the additional set of canonical equations
\begin{equation}\label{ErwKanGl2}
\pfrac{\tilde{t}\indices{_\mu^\nu}}{X^\nu} = \left.\pfrac{\tilde{\mathcal{H}}_e}{x^\mu}\right|_\text{expl} \quad \text{and} \quad \pfrac{x^\mu}{X^\nu} = -\pfrac{\tilde{\mathcal{H}}_e}{\tilde{t}\indices{_\mu^\nu}}.
\end{equation}
Thereby, the dynamics is equivalent to that of the unparametrized system.

There is another extra condition in the extended Hamiltonian formalism. The reparametrization invariance of the extended Lagrangian $\tilde{\mathcal{L}}_e$ at the same time means that it is homogeneous in all derivatives with respect to each 1 of the new parameters $X^\mu$. In this case \textbf{Euler's theorem on homogeneous functions} (see e.~g. \cite{Lanczos}) says that there are the identities
\begin{equation}
\pfrac{\tilde{\mathcal{L}}_e}{\pfrac{\phi}{X^\mu}} \pfrac{\phi}{X^\mu} + \pfrac{\tilde{\mathcal{L}}_e}{\pfrac{x^\alpha}{X^\mu}} \pfrac{x^\alpha}{X^\mu} \equiv \tilde{\mathcal{L}}_e \quad \forall \mu=0,\ldots,m-1.
\end{equation}
In point mechanics the analogous relation would force the Hamiltonian to vanish \cite{ExtPointMechanics}, but in field theory the setup is a bit different. It is now a constraint on the new momentum variables $\tilde{t}\indices{_\mu^\nu}$. As it is an identity, the condition remains valid upon taking derivatives. So take the derivative of this identity with respect to the metric $(g_{\mu\nu})$ for the tensor relation
\begin{equation}\label{HomBed}
\pfrac{\tilde{\mathcal{L}}_e}{\pfrac{\phi}{X^\nu}} \pfrac{\phi}{X^\mu} + \pfrac{\tilde{\mathcal{L}}_e}{\pfrac{x^\alpha}{X^\nu}} \pfrac{x^\alpha}{X^\mu} \equiv \tilde{\pi}^\nu \pfrac{\phi}{X^\mu} - \tilde{t}\indices{_\alpha^\nu} \pfrac{x^\alpha}{X^\mu} = \delta_\mu^\nu \tilde{\mathcal{L}}_e \quad \forall \mu,\nu=0,\ldots,m-1.
\end{equation}
With simple recasting one can extract the energy-momentum tensor via
\begin{equation}
\begin{aligned}
\tilde{t}\indices{_\mu^\nu} &:= -\pfrac{\tilde{\mathcal{L}}_e}{\pfrac{x^\mu}{X^\nu}} \equiv \pfrac{\tilde{\mathcal{L}}_e}{\pfrac{\phi}{X^\nu}} \pfrac{\phi}{X^\beta} \pfrac{X^\beta}{x^\mu} - \delta_\beta^\nu \tilde{\mathcal{L}}_e \pfrac{X^\beta}{x^\mu} \\
&\equiv \theta\indices{_\beta^\nu}(X) \pfrac{X^\beta}{x^\mu} \det \pfrac{x}{X} \equiv \theta\indices{_\mu^\alpha}(x) \pfrac{X^\nu}{x^\alpha} \det \pfrac{x}{X}.
\end{aligned}
\end{equation}
As an identity it holds on the full extended configuration space. Thereby, the fundamental role of the energy-momentum tensor is strengthened again as the counterpart to the mechanics Hamiltonian / energy. This is especially interesting with regard to gravity where this identity is already the whole expression of the Einstein field equations and whose theory explicitly builds on covariance.

\subsection{Correction of the extended Poisson brackets}
In a straightforward manner the \textbf{extended Poisson brackets} are defined as
\begin{equation}
\{F, G\}_{e,\phi, \tilde{\pi}^\nu, x^\mu, \tilde{t}\indices{_\mu^\nu}} := \pfrac{F}{\phi} \pfrac{G}{\tilde{\pi}^\nu} - \pfrac{F}{\tilde{\pi}^\nu} \pfrac{G}{\phi} - \pfrac{F}{x^\mu} \pfrac{G}{\tilde{t}\indices{_\mu^\nu}} + \pfrac{F}{\tilde{t}\indices{_\mu^\nu}} \pfrac{G}{x^\mu}.
\end{equation}

The quantity to look for when regularizing the bracket of an arbitrary function $F$ is, according to the analysis before, the spacetime conjugate momentum $\tilde{t}\indices{_\mu^\nu}$. The additional set of canonical equations (\ref{ErwKanGl2}) guarantees that still only its divergence is determined by the extended Hamiltonian system. Then, in analogy to the simply covariant case, one may define
\begin{equation}\label{H2}
\tilde{h}\indices{_\nu^\mu} := \pfrac{}{X^\beta} \left(\phi \left(\delta_\nu^\beta \tilde{\pi}^\mu - \delta_\nu^\mu \tilde{\pi}^\beta\right)\right).
\end{equation}
This tensor is also itself divergence-free, see the appendix. This means that addition or subtraction of it does not affect the extended energy-momentum tensor. So this gives rise to the \textbf{equivalent tensor}
\begin{equation}
\tilde{T}\indices{_\nu^\mu} := \pfrac{x^\beta}{X^\nu} \tilde{t}\indices{_\beta^\mu} - \tilde{h}\indices{_\nu^\mu}.
\end{equation}
Using the homogeneity condition (\ref{HomBed}) it can as well be expressed as
\begin{equation}
\tilde{T}\indices{_\nu^\mu} = -\delta_\nu^\mu \tilde{\mathcal{L}}_e + \tilde{\pi}^\mu \pfrac{\phi}{X^\nu} - \pfrac{}{X^\beta} \left(\phi \left(\delta_\nu^\beta \tilde{\pi}^\mu - \delta_\nu^\mu \tilde{\pi}^\beta\right)\right).
\end{equation}
Note that strictly speaking the Lagrangian is a function on configuration space instead of phase space, but it has to remain in a mixed form because the energy-momentum tensor itself is translated as a canonical momentum. According to the homogeneity condition (\ref{HomBed}), there is an identity, so this representation does not harm the following calculations and the expression can be used safely.

There are now analogous field equations to (\ref{ThetaKanGl}):
\begin{equation}
\pfrac{\tilde{T}\indices{_\nu^\mu}}{\phi} = -\delta_\nu^\mu \pfrac{\tilde{\mathcal{L}}_e}{\phi} - \pfrac{\tilde{\pi}^\mu}{X^\nu} + \delta_\nu^\mu \pfrac{\tilde{\pi}^\beta}{X^\beta} = -\pfrac{\tilde{\pi}^\mu}{X^\nu}
\end{equation}
and
\begin{equation}
\pfrac{\tilde{T}\indices{_\nu^\mu}}{\tilde{\pi}^\mu} = \pfrac{\phi}{X^\nu} - \pfrac{\phi}{X^\nu} + \delta_\nu^\mu \pfrac{\phi}{X^\beta} \delta_\mu^\beta = \pfrac{\phi}{X^\nu}.
\end{equation}
Moreover, the additional set of canonical equations (\ref{ErwKanGl2}) has a counterpart as well:
\begin{equation}
\pfrac{\tilde{T}\indices{_\nu^\mu}}{x^\alpha} = \pfrac{x^\beta}{X^\nu} \pfrac{\tilde{t}\indices{_\beta^\mu}}{x^\alpha} \quad \text{and} \quad \pfrac{\tilde{T}\indices{_\nu^\mu}}{\tilde{t}\indices{_\beta^\mu}} = \pfrac{x^\beta}{X^\nu}.
\end{equation}
They determine all the partial derivatives of the momenta, hence are richer than the canonical equations. They can finally be used to calculate the extended Poisson bracket of an arbitrary function $F$ with that tensor:
\begin{equation}\label{ExtPoisHeisenberg}
\begin{aligned}
\{F, \tilde{T}\indices{_\nu^\mu}\}_{e,\mu} &= \pfrac{F}{\phi} \pfrac{\tilde{T}\indices{_\nu^\mu}}{\tilde{\pi}^\mu} - \pfrac{F}{\tilde{\pi}^\mu} \pfrac{\tilde{T}\indices{_\nu^\mu}}{\phi} + \pfrac{F}{x^\beta} \pfrac{\tilde{T}\indices{_\nu^\mu}}{\tilde{t}\indices{_\beta^\mu}} - \pfrac{F}{\tilde{t}\indices{_\beta^\mu}} \pfrac{\tilde{T}\indices{_\nu^\mu}}{x^\beta} \\
&= \pfrac{F}{\phi} \pfrac{\phi}{X^\nu} + \pfrac{F}{\tilde{\pi}^\mu} \pfrac{\tilde{\pi}^\mu}{X^\nu} + \pfrac{F}{x^\beta} \pfrac{x^\beta}{X^\nu} - 0 \\
&= \pfrac{F}{X^\nu}.
\end{aligned}
\end{equation}
In the end, the goal of a bracket relation that returns the total differential of the function $F$ is achieved.

\section{Outlook: Quantization}
\subsection{Field theoretic quantization of spacetime}
The relation (\ref{ExtPoisHeisenberg}) is especially appealing for a field theoretic quantization. There is the general question how to transfer a vectorial Poisson bracket $\{\cdot,\cdot\}_\mu$ to a scalar commutator $[\cdot,\cdot]$. For the classical fields this problem is solved by slicing the temporal component (bosons) resp. taking the trace with respect to the temporal gamma matrix (fermions). But there is one further theory that lacks a proper quantization at all and where covariance is a fundamental principle: gravitation. The quantity appearing in eq. (\ref{ExtPoisHeisenberg}) is just an equivalent representation of the energy-momentum tensor, the source of gravity. So it may be worth considering this one part of a fundamental canonical commutator relation.

Another consideration is that general relativity is a theory that describes the curvature of spacetime itself. So instead of vainly trying to quantize the metric or the connection, the answer may lie in quantizing spacetime itself. There are models that do that, most prominently \textbf{noncommutative spacetime} \cite{NonCommGeo, YangNCG}. But they essentially discuss commutator relations between spacetime coordinates themselves.

The admittedly speculative idea is to combine both approaches in recognizing that in the parametrized theory from above those are a canonical variable pair. The general rationale from quantization tells that they should obey a ``fundamental commutator relation''. In this context one attempt may be
\begin{equation}\label{RaumzeitKomm}
\begin{aligned}[]
[\hat{x}_\mu(X), \hat{x}_\nu(X')] &\overset{!}{=} 0 \\
[\hat{\tilde{t}}\indices{_\mu^\alpha}(X), \hat{\tilde{t}}\indices{_\nu^\beta}(X')] &\overset{!}{=} 0 \\
[\hat{x}_\mu(X), -\hat{\tilde{t}}\indices{_\nu^\alpha}(X')] &\overset{!}{=} i\hbar \widehat{\pfrac{x_\mu}{X^\nu}} \widehat{\pfrac{X^\alpha}{x^\mu}} \det \widehat{\pfrac{x}{X}} \delta^{(4)}(X-X').
\end{aligned}
\end{equation}
This is a proper field theoretic analog to usual commutators. The coordinate expression may be understood with a fixed gauge, as is common practice in quantum field theory. The main specialty is, however, a $\delta$ distribution on the whole parameter space. This choice prevents particle propagation processes. On the one hand, this ensures that spacetime does not lose its meaning as the frame on which the physical objects live. On the other hand, this gives rise to a bit different treatment.

Note that $\tilde{t}\indices{_\mu^\alpha}$ denotes the whole Einstein field equation complex, i.~e. the energy-momentum tensor plus the Einstein tensor. The usual issue for quantum gravity that only one side of the Einstein field equations would be an operator and the other side a simple function is thus directly done. However, there is the well-known problem from gauge theories that a valid Einstein field equation spoils the commutator relations (\ref{RaumzeitKomm}). So there must be some gauge-fixing term to be added to the quantum Lagrangian. A minimal correction that serves the desired purpose is
\begin{equation}\label{GFRaumzeit}
\tilde{\mathcal{L}}_{GF} = -C \det \pfrac{x}{X}
\end{equation}
for a constant $C$. Splitting it into $C = \kappa \cdot \Lambda$ (such that $\kappa$ is the internal coupling constant between gravity and energy-momentum), this can be interpreted as the inclusion of a \textbf{cosmological constant} $\Lambda$, whose value does not have to equal a huge amount, but should literally be ``minimal''.

A gauge theory further necessitates a physicality condition to rule out which configurations are realized in nature. This is generally mediated by a conserved quantity, which is that tensor according to the extended canonical equation (\ref{ErwKanGl2}). A reasonable choice would then be
\begin{equation}\label{RZPhys}
\hat{\tilde{t}}\indices{_\nu^\mu}|\psi_\text{phys}\rangle \overset{!}{=} \langle\psi_\text{phys}|\hat{\tilde{t}}\indices{_\nu^\mu} \overset{!}{=} 0,
\end{equation}
which determines the value of the whole Einstein term and says nothing else than that in the physical subspace the Einstein field equations are valid. Thereby, it is not a dynamical equation any more, but a constraint.

Another interesting point is the \textbf{uncertainty relations}. They are derived via a general operator identity. Application of that formula here yields:
\begin{equation}
\Delta \hat{x}_\mu(X) \Delta \hat{\tilde{t}}\indices{_\nu^\alpha}(X) \geq \frac{1}{2} |\langle\psi| [\hat{x}_\mu(X), \hat{\tilde{t}}\indices{_\nu^\alpha}(X)] |\psi\rangle| = \frac{\hbar}{2} \left|\widehat{\pfrac{x_\mu}{X^\nu}} \widehat{\pfrac{X^\alpha}{x^\mu}}\right|.
\end{equation}
In the special case of the trace of the spacetime momentum tensor this gives
\begin{equation}\label{RaumzeitUnschaerfe}
\Delta \hat{x}_\mu(X) \Delta \hat{\tilde{t}}\indices{_\nu^\nu}(X) \geq \frac{\hbar}{2}.
\end{equation}
The interpretation is that spacetime and the gravitational action of energy and matter are not simultaneously perfectly measurable, in the same way how space and momentum or time and energy are not simultaneously sharp in quantum mechanics. Especially, the energy-momentum tensor cannot become exactly traceless and spacetime not vanishing, which gives a formal barrier to a \textbf{big bang}.

\subsection{Similarity to the BRST formalism}
With the parametrized formalism another remarkable similarity can be drawn, namely to \textbf{BRST quantization} \cite{BRS, Tyutin}. For the undertaken comparison consider the quantum electrodynamics Lagrangian
\begin{equation}
\mathcal{L}_B = \mathcal{L}_{EM} + B \partial^\mu A_\mu + \frac{1}{2} \alpha B^2.
\end{equation}
$\mathcal{L}_{EM}$ denotes the classical electrodynamics (Maxwell) Lagrangian, but the focus here lies on the gauge-fixing term that appears in covariant formulations of quantum gauge theories because of the redundant gauge orbits. $\alpha$ is commonly called the gauge parameter and $B(x)$ is a (Nakanishi-Lautrup) multiplier field. Since $\mathcal{L}_{EM}$ does not depend on $B$, variation with respect to this field yields
\begin{equation}\label{GlBFeld}
\partial^\mu A_\mu + \alpha B = 0.
\end{equation}
The next interesting quantity is the covariant Hamiltonian. Therefore, a partial Legendre transformation with respect to the physical fields has to be employed as the gauge-fixing term is singular, hence non-transformable. This essentially results in a sign flip for the gauge-fixing term:
\begin{equation}
\mathcal{H}_B = \mathcal{H}_{EM} - B \partial^\mu A_\mu - \frac{1}{2} \alpha B^2.
\end{equation}

This can now be compared to the extended Hamiltonian. It acquires an extra term consisting of the new momentum field $\tilde{t}\indices{_\nu^\mu}$ due to the reparametrization as well as the gauge fixing term (\ref{GFRaumzeit}) from the quantization ansatz above:
\begin{equation}
\tilde{\mathcal{H}}_e = \tilde{\mathcal{H}}_0 - \tilde{t}\indices{_\nu^\mu} \pfrac{x^\nu}{X^\mu} + \kappa\Lambda \det \pfrac{x}{X}.
\end{equation}
There is a remarkable similarity: Taking into account the simple identity $\pfrac{x^\nu}{X^\mu} \pfrac{X^\mu}{x^\nu} = 4,$ one may interpret the term $-\frac{1}{2} \kappa \Lambda$ as the \textbf{gauge parameter} ($\alpha$), $\pfrac{x^\nu}{X^\mu}$ corresponds to the ``$B$''-\textbf{field}, and the remaining contribution $t\indices{_\nu^\mu}$ is the \textbf{gauge-fixing condition}. This last similarity is what was exactly exploited in (\ref{RZPhys}). Upon comparison with non-Abelian gauge theories, it can readily be observed that the parametrized formalism is free of ghosts. The field equation corresponding to the auxiliary field dynamics (\ref{GlBFeld}) here becomes
\begin{equation}
\tilde{t}\indices{_\nu^\mu} - \frac{1}{2} \kappa\Lambda \pfrac{X^\mu}{x^\nu} \det \pfrac{x}{X} = 0.
\end{equation}
Another analogy can be drawn by looking at the definition (\ref{RaumzeitImpuls}) of these new momenta. Define the quantity
\begin{equation}
\tilde{j}\indices{_\mu^\nu} := \pfrac{x^\alpha}{X^\mu} \pfrac{\tilde{\mathcal{L}}_e}{\pfrac{x^\alpha}{X^\nu}} = - \pfrac{x^\alpha}{X^\mu} \tilde{t}\indices{_\alpha^\nu}.
\end{equation}
The symbolism shall reflect the similarity to a \textbf{BRST current}. Normally, it is constructed by an infinitesimal symmetry transformation and the physicality condition then uses the space-integral of the temporal component (BRST charge). That it is here a finite transformation and the condition (\ref{RZPhys}) is set up covariantly, appears even cleaner and perfectly fits a covariant theory like gravity. All in all, there are remarkable similarities to BRST quantization. In contrast, this formalism is intrinsic in its nature, well-defined in the Hamiltonian regime and physical in the large-scale dynamics of the universe. It would be interesting if a formal translation could be established and the quantization approach outlined here has a physical significance.

\appendix
\section{Explicit calculation of certain equations}
\paragraph{Eq. (\ref{PoissonDeviation})}
The total derivative of a function $F(\phi, \pi^\mu, x^\mu)$ on phase space can be decomposed as $$\pfrac{F}{x^\mu} = \pfrac{F}{\phi} \pfrac{\phi}{x^\mu} + \pfrac{F}{\pi^\nu} \pfrac{\pi^\nu}{x^\mu} + \left.\pfrac{F}{x^\mu}\right|_\text{expl}.$$ Using this and the canonical equations (\ref{KanGl}) for the De~Donder-Weyl Hamiltonian $\mathcal{H}$, the Poisson bracket becomes
\begin{align*}
\{F,\mathcal{H}\}_\mu &= \pfrac{F}{\phi} \pfrac{\mathcal{H}}{\pi^\mu} - \pfrac{F}{\pi^\mu} \pfrac{\mathcal{H}}{\phi} \\
&= \pfrac{F}{\phi} \pfrac{\phi}{x^\mu} + \pfrac{F}{\pi^\mu} \pfrac{\pi^\nu}{x^\nu} \\
&= \pfrac{F}{x^\mu} - \left.\pfrac{F}{x^\mu}\right|_\text{expl} + \pfrac{F}{\pi^\mu} \pfrac{\pi^\nu}{x^\nu} - \pfrac{F}{\pi^\nu} \pfrac{\pi^\nu}{x^\mu}.
\end{align*}

\paragraph{Eq. (\ref{DivEnImp})}
One may use the Hamiltonian representation of the energy-momentum tensor (\ref{EnImpTensor}) and then insert the canonical equations (\ref{KanGl}):
\begin{align*}
\pfrac{\theta\indices{_\nu^\mu}}{x^\mu} &= \pfrac{\pi^\mu}{x^\mu} \pfrac{\mathcal{H}}{\pi^\nu} + \pi^\mu \pfrac{}{x^\mu} \pfrac{\mathcal{H}}{\pi^\nu} \\
&\quad + \delta_\nu^\mu \left(\pfrac{\mathcal{H}}{\phi} \pfrac{\phi}{x^\mu} + \pfrac{\mathcal{H}}{\pi^\beta} \pfrac{\pi^\beta}{x^\mu} + \left.\pfrac{\mathcal{H}}{x^\mu}\right|_\text{expl} - \pfrac{\pi^\beta}{x^\mu} \pfrac{\mathcal{H}}{\pi^\beta} - \pi^\beta \pfrac{}{x^\mu} \pfrac{\mathcal{H}}{\pi^\beta}\right) \\
&= -\pfrac{\mathcal{H}}{\phi} \pfrac{\phi}{x^\nu} + \pi^\mu \pfrac{}{x^\mu} \pfrac{\phi}{x^\nu} \\
&\quad + \delta_\nu^\mu \left(\pfrac{\mathcal{H}}{\phi} \pfrac{\phi}{x^\mu} + \pfrac{\mathcal{H}}{\pi^\beta} \pfrac{\pi^\beta}{x^\mu} + \left.\pfrac{\mathcal{H}}{x^\mu}\right|_\text{expl} - \pfrac{\pi^\beta}{x^\mu} \pfrac{\mathcal{H}}{\pi^\beta} - \pi^\beta \pfrac{}{x^\mu} \pfrac{\phi}{x^\beta}\right) \\
&= \left.\pfrac{\mathcal{H}}{x^\nu}\right|_\text{expl}.
\end{align*}

\paragraph{Vanishing divergence of (\ref{ModTheta})}
It only needs to be shown that the divergence of $$h\indices{_\nu^\mu} := \pfrac{}{x^\beta} \left(\phi \left(\delta_\nu^\beta \pi^\mu - \delta_\nu^\mu \pi^\beta\right)\right)$$ vanishes. This can be done easily:
\begin{align*}
\pfrac{h\indices{_\nu^\mu}}{x^\mu} &= \ppfrac{\phi}{x^\mu}{x^\nu} \pi^\mu + \pfrac{\phi}{x^\mu} \pfrac{\pi^\mu}{x^\nu} + \pfrac{\phi}{x^\nu} \pfrac{\pi^\mu}{x^\mu} + \phi \ppfrac{\pi^\mu}{x^\mu}{x^\nu} \\
&\quad - \ppfrac{\phi}{x^\nu}{x^\beta} \pi^\beta - \pfrac{\phi}{x^\nu} \pfrac{\pi^\beta}{x^\beta} - \pfrac{\phi}{x^\beta} \pfrac{\pi^\beta}{x^\nu} - \phi \ppfrac{\pi^\beta}{x^\nu}{x^\beta} \\
&\equiv 0.
\end{align*}

\paragraph{Vanishing divergence of (\ref{H2})}
The calculation goes the same way like before:
\begin{align*}
\pfrac{\tilde{h}\indices{_\nu^\mu}}{X^\mu} &= \ppfrac{\phi}{X^\mu}{X^\nu} \tilde{\pi}^\mu + \pfrac{\phi}{X^\mu} \pfrac{\tilde{\pi}^\mu}{X^\nu} + \pfrac{\phi}{X^\nu} \pfrac{\tilde{\pi}^\mu}{X^\mu} + \phi \ppfrac{\tilde{\pi}^\mu}{X^\mu}{X^\nu} \\
&\quad - \ppfrac{\phi}{X^\nu}{X^\beta} \tilde{\pi}^\beta - \pfrac{\phi}{X^\nu} \pfrac{\tilde{\pi}^\beta}{X^\beta} - \pfrac{\phi}{X^\beta} \pfrac{\tilde{\pi}^\beta}{X^\nu} - \phi \ppfrac{\tilde{\pi}^\beta}{X^\nu}{X^\beta} \\
&\equiv 0.
\end{align*}


\begin{thebibliography}{99}

\bibitem{Redelbach}
J. Struckmeier and A. Redelbach (2008): \emph{Covariant Hamitonian Field Theory}, Int. J. Mod. Phys. E 17 (3), 435-491

\bibitem{DiracQuant}
P. A. M. Dirac (1958): \emph{The Principles of Quantum Mechanics}, 4th ed., Oxford Science Publications, ISBN 978-0-19-852011-5

\bibitem{Kanatchikov}
I. V. Kanatchikov (1997): \emph{On Field Theoretic Generalizations of a Poisson Algebra}, Rept. Math. Phys. 40 (2), 225-234

\bibitem{KanatchikovQuant}
I. V. Kanatchikov (2001): \emph{Precanonical Quantum Gravity: Quantization Without the Space-Time Decomposition}, Int. J. Theor. Phys. 40 (6), 1121-1149

\bibitem{Paufler}
M. Forger, C. Paufler and H. Römer (2003): \emph{The Poisson Bracket for Poisson Forms in Multisymplectic Field Theory}, Rev. Math. Phys. 15 (7), 705-743

\bibitem{HippelWohlfarth}
G. M. von Hippel and M. N. R. Wohlfarth (2006): \emph{Covariant canonical quantization}, Eur. Phys. J. C 47, 861-872

\bibitem{GRextended}
J. Struckmeier (2015): \emph{General relativity as an extended canonical gauge theory}, Phys. Rev. D 91, 085030

\bibitem{ParamField}
M. Castrillón López and M. J. Gotay (2010): \emph{Covariantizing Classical Field Theories}, arXiv:1008.3170

\bibitem{ExtPointMechanics}
J. Struckmeier (2009): \emph{Extended Hamilton-Lagrange formalism and its application to Feynman's path integral for relativistic quantum physics}, Int. J. of Mod. Phys. E18, 79-108

\bibitem{Lanczos}
C. Lanczos (1970): \emph{The Variational Principles of Mechanics}, 4th ed., University of Toronto Press, ISBN 0-486-65067-7

\bibitem{NonCommGeo}
H. S. Snyder (1947): \emph{Quantized Space-Time}, Phys. Rev. 71 (1), 38

\bibitem{YangNCG}
C. N. Yang (1947): \emph{On Quantized Space-Time}, Phys. Rev. 72 (9), 874

\bibitem{BRS}
C. Becchi, A. Rouet and R. Stora (1976): \emph{Renormalization of gauge theories}, Annals of Physics 98 (2), 287–321

\bibitem{Tyutin}
I. V. Tyutin (1975): \emph{Gauge Invariance in Field Theory and Statistical Physics in Operator Formalism}, Lebedev Physics Institute preprint 39

\end{thebibliography}
\end{document}